\documentclass{JHEP3}
\def\beq{\begin{equation}}
\def\endeq{\end{equation}}
\def\bea{\begin{eqnarray}}
\def\endea{\end{eqnarray}}
\def\O{{\cal{O}}}
\def\dsl{{\rlap{\kern0.5pt /}{\partial}}}

\def\lto{\mathop
        {\hbox{${\lower3.8pt\hbox{$<$}}\atop{\raise0.2pt\hbox{$\sim$}}$}}}

\title{Quantum Polarization of D4-branes}

\author{Belkis Cabrera Palmer\\
Physics Department, Syracuse University, Syracuse, NY 13244,\\
Physics Department, UCSB, Santa Barbara, CA 93106. \texttt{bcabrera@physics.ucsb.edu}}

\author{Donald Marolf\\
Physics Department, UCSB, Santa Barbara, CA 93106.
\texttt{marolf@physics.ucsb.edu}}

\author{Pedro J. Silva\\
Dipartimento di Fisica dellÕUniversit\'a di Milano, ViaCeloria16, I-20133 Milano, Italy \\
INFN, Sezione di Milano, Via Celoria16, I-20133 Milano, Italy
\texttt{pedro.silva@mi.infn.it}}

\abstract{
The low energy effective field theory of type II D4-branes coupled to bulk supergravity fields is used to investigate {\it quantum} effects for D4-branes in the D0 supergravity background.  Classically, the D4-branes are unaffected by this background.  However, quantum (one-loop) effects are argued to lead to an induced density of D0-brane charge; e.g., D0-multipole moments on the D4-brane.  The effect is divergent in field theory, but is expected to be cut-off naturally by stringy corrections.  }

\date{August, 2004}

\keywords{Brane Polarization, D-branes}

\begin{document}

\section{Introduction}

The polarization of branes by external fields \cite {pol1,pol2} is an important effect with many applications in string theory including those associated with giant gravitons \cite{gg1}, instanton studies \cite{inst}, brane inflation \cite{henry},  and supertubes \cite{st1,st2}, the last of these  being the subject of an interesting proposal \cite{LuninMathur1,LuninMathur2,MathurEssay} regarding black hole entropy.  To our knowledge, however, all of the standard applications can be viewed as {\it classical} polarization effects, perhaps with quantum corrections.  Consider, however, $Dp$-branes in the supergravity fields generated by $D(p\pm4)$-branes.  In this context the unpolarized $Dp$-branes are known to saturate the BPS bound.  Thus, there can be no lower energy classical configuration and the classical ground state remains unpolarized\footnote{There could, however, be several degenerate ground states so long as one remains unpolarized.}.  However, this argument leaves open the possible distortion of quantum fluctuations around the classical ground state and associated {\it quantum} polarization effects.

Such quantum polarizations were first explored in \cite{MS}, which considered a bound state of $N$ test $D0$-branes placed in the supergravity background generated by a collection of parallel $D4$-branes.  It was argued under such conditions that, to lowest order in the weak field limit, the width of the $D0$-bound state changes by an amount proportional to $R_0 (gN)^{1/3} f^2$, where $R_0$ is the unperturbed width of the bound state and $f$ is a dimensionless measure of the Ramond-Ramond field strength at the $D0$-branes.  This result was shown to match the corresponding distortion of the near-$D0$ supergravity solution in the manner expected from gauge/gravity duality.

Here we study the opposite limit and consider a test $D4$-brane placed in the supergravity background generated by a collection of $D0$-branes.  The advantage of this context is that one may find effects even for the abelian theory associated with a single brane.  Instead of examining the size of the brane, we compute the induced density of $D0$-brane charge, $\langle \rho_{D0} \rangle$, defined by the coupling of the $D4$-brane to the Ramond-Ramond 1-form $C^{(1)}$.  In the approximation of interest, this charge density is proportional to $F \wedge F$ plus a quadratic fermion term, where $F$ is the (abelian) Yang-Mills fields on the $D4$-brane.  Because $\rho_{D0}$ contains quadratic terms, the expectation value $\langle \rho_{D0} \rangle$ is sensitive to quantum fluctuations.  Note, however, that this is indeed a polarization effect as the integral of $\rho_{D0}$ must vanish.

The calculations below are performed using the low energy effective field theory for the $D4$-brane, including the couplings of world-volume Fermions to bosonic supergravity backgrounds found in \cite{MMS1,MMS2}.  We find that $\langle \rho_{D0}\rangle$ does not vanish in the supergravity background generated by $D0$-branes.  Instead, it diverges in our field theory treatment.  This is somewhat surprising given the supersymmetry of our setting\footnote{In particular, as shown in \cite{MMS2} our field theory retains an explicit invariance under an 8-supercharge supersymmetry algebra when coupled to the $D0$ background.}, but appears not to contradict any known results.  In a full string-theoretic treatment one naturally expects that this divergence will be cutoff at the string scale.

We begin in section \ref{prelim} below with a short review of the results of \cite{MMS2} and a precise statement of our setup.  The field-theoretic calculation of $\langle \rho_{D0} \rangle$ is then presented in section \ref{calc}.  As is clear from the above description, our calculation begs a full string-theoretic treatment.  While we defer such a calculation to future work, it is interesting to assume that a full string treatment cuts off our divergences at the string scale but leaves them non-vanishing, and to consider the implications.  We discuss such implications in section \ref{disc}, showing that such a term has the right form to arise from a 1-loop (annulus) string diagram.  In particular, our polarization effect would require the $D4$-brane effective action to have a 1-loop term of the form $\int d^5 x |p(dC^{(1)})|^2$, where
$p(dC^{(1)})$ denotes the pull-back of  the bulk 2-form field strength $dC$ to the brane and the notation
$|p(dC^{(1)})|^2 = [p(dC^{(1)})]^{IJ} [p(dC^{(1)})]_{IJ}$, where the contraction is performed using the induced metric on the brane.    We will use $I,J$ to denote worldvolume indices and $A,B$ to denote spacetime directions.  It is convenient to mention here that we use analogous notation $\hat I, \hat J$ and $\hat A, \hat B$ for tangent space directions, and similarly $i,j$ and $a,b$ for world-volume and spacetime {\it spatial} directions (i.e., orthogonal to the $D0$ worldlines) and $\hat i, \hat j, \hat a, \hat b$ for the corresponding tangent space directions.

\section{Preliminaries}
\label{prelim}

Recall that our goal is to study deformations of the $D4$-brane ground state when placed in the supergravity field generated by a collection of $D0$-branes.  We shall therefore take the $D4$-brane as a test object whose back-reaction on the supergravity fields can be ignored.  One expects this approximation to be valid in the limit where the string coupling $g$ is taken to zero but the number of $D0$-branes is increased so that the supergravity background remains fixed.

\subsection{The charge density operator, $\rho_{D0}$}

Since we wish to compute the charge density which couples to the Ramond-Ramond vector potential $C_A$, and since this charge is defined by varying the $D4$-brane action with respect to $C_A$, we will need the general coupling of the $D4$-brane to this field.  The coupling of the bosonic $D4$-brane fields is familiar, but the Fermion couplings are more complicated.  The complete set of such couplings was calculated in \cite{MMS2} to quadratic order in Fermions.  This will suffice for our purposes as we have already stated that we will take $g$ small,  and so may work perturbatively in world-volume fields.

The lowest order effect is thus given by the quadratic truncation of the $D4$-brane effective action, which is just the $N=4$ $U(1)$ theory coupled to our background (\ref{back}).  In particular, the Fermion terms we require will be second order in Fermions and will involve no coupling to the world-volume gauge field $F_{IJ}$.  Thus we will use a truncated effective action of the form
\begin{eqnarray}
\label{Dpt}
S_{D4}^{trunc}&=&S^{(0)}_{D4}+S^{(2) \ trunc}_{D4}, \cr
S_{D4}^{(0)}&=& -T_{D4}\int d^5\xi e^{-\phi}\sqrt{-(g+{\cal
F})}+T_{D4}\int C\; e^{-{\cal F}}\,,
\end{eqnarray}
where $S^{(2) \ trunc}_{Dp}$ will contain the appropriate quadratic Fermion terms, $g$ is the induced metric, ${\cal F}_{IJ} = F_{IJ} + B_{IJ}$, $C = \sum_n C^{(n)}$ is a formal sum of the IIA Ramond-Ramond potentials and the integral $\int C\; e^{-{\cal F}}$ picks out the form of rank $5$ to integrate.   We will also use $X^A(\xi)$ to denote the embedding of the brane in spacetime.

The quadratic Fermion term is written in terms of a real Majorana Fermion $\psi$, which lives in the 32-component representation of the Clifford algebra
\begin{equation}
\{\Gamma^{\hat A},\Gamma^{\hat B}\} = 2 \eta^{\hat A \hat B}.
\end{equation}
The conjugate spinor $\bar \psi$ is defined by $\bar \psi_\beta = \psi^\alpha C_{\alpha \beta}$, where $C$ is the anti-symmetric charge-conjugation matrix which we take to be $C_{\alpha \beta } \equiv \Gamma^{\hat 0 \beta}{}_\alpha$.
Following \cite{MMS2}, we use of the notation $\Gamma^{\hat \varphi} =  \Gamma^{\hat 0}  \Gamma^{\hat 1} \Gamma^{\hat 2} \Gamma^{\hat 3} \Gamma^{\hat 4} \Gamma^{\hat 5} \Gamma^{\hat 6} \Gamma^{\hat 7} \Gamma^{\hat 8} \Gamma^{\hat 9}$ for the ten-dimensional chirality operator.  We also use the notation
$\Gamma_{D4}={1\over 5!\sqrt{-g}}\tilde \epsilon^{IJKLM}\Gamma_{IJKLM}\Gamma^{\hat \varphi}$ for an interesting world-volume chirality operator, where $\tilde \epsilon$ denotes the Levi-Civita tensor density (which takes value $\pm1,0$ for any metric).  Finally, we will use the notation
$\Gamma_{I_1...I_n}=\Gamma_{[I_1}...\Gamma_{I_n]}$
denoting antisymmetrization with weight one; e.g. $\Gamma_{01} = \hbox{${1\over2}$}(\Gamma_0 \Gamma_1- \Gamma_1 \Gamma_0) = \Gamma_0 \Gamma_1$.

We may also drop any couplings of Fermions to the background Neveu-Schwarz two-form $B_{AB}$ (though these are non-trivial and were computed in \cite{MMS1,MMS2}) since it will vanish in the background generated by $D0$-branes and we will not need to vary it.
With this understanding the truncated quadratic Fermion action $S^{(2)\ trunc}_{D4}$ may be seen from \cite{MMS2} to be
\begin{equation}
S^{(2) \  trunc}_{D4}=\frac
{iT_{D4}}{2}\int d^5\xi e^{-\phi}\sqrt{-g}\bar \psi(1-\Gamma_{D_4})(\Gamma^ID_I-\Delta)\psi ,
\end{equation}
where
\begin{eqnarray}
D_A &=& \partial_A +\frac{1}{4} \omega_{A\hat B \hat C}\Gamma^{\hat B \hat C} +
\frac18 e^\phi \left( \frac{1}{2!} {\bf F}^{(2)}_{\hat B \hat C}\Gamma^{\hat B \hat C }\Gamma_A\Gamma^{\hat {\varphi}}+
\frac{1}{4!}{\bf F}^{(4)}_{\hat B \hat C \hat D \hat E}\Gamma^{\hat B \hat C \hat D \hat E}\Gamma_A\right) \ {\rm and} \cr
\Delta &=& \frac{1}{2} \Gamma^A \partial_A \phi +
\frac{1}{8} e^\phi \left( \frac{3}{2!} {\bf F}^{(2)}_{\hat B \hat C}\Gamma^{\hat B \hat C}\Gamma^{\hat {\varphi}}+
\frac{1}{4!} {\bf F}^{(4)}_{\hat B \hat C \hat D \hat E}\Gamma^{\hat B \hat C \hat D \hat E}\right)\ .
\end{eqnarray}
Here $\omega$ is the spin connection of the spacetime metric and we have chosen to denote bulk Ramond-Ramond fields by bold-face ${\bf F}^{(n)}= dC^{(n-1)} + {Wess-Zumino} \ terms$ in order to distinguish them from the $U(1)$ world-volume field $F$ on the $D4$-brane.  The superscript $(n)$ denotes the rank of the form.

It is now straightforward to vary the action (\ref{Dpt}) and obtain the current $J^A \equiv \frac{1}{\sqrt{-g}} \frac{\delta S_{D4}^{trunc}}{\delta C_A^{(1)}}$ that couples to $C_A^{(1)}$.  The result is
\begin{equation}
\label{current}
J^A = \frac{T_{D4}}  {8\sqrt{-g}}   \frac{\partial X^A}{\partial \xi^M} \tilde \epsilon^{MIJKL} F_{IJ} F_{KL} +i \frac{T_{D4}}{8\sqrt{-g}} \partial_B \left( \sqrt{-g} \bar \psi (1 -  \Gamma_{D4}) \left(- \Gamma^{BA}    +2 \Gamma^I\partial_I(X^{[B})\Gamma^{A]} \right) \Gamma^{\hat \varphi} \psi \right).
\end{equation}
We will in particular be interested in the charge density $\rho_{D0} \equiv J^0$, where $0$ denotes the direction along the world-lines of the $D0$-branes that generate the background of interest, as all other components of $J^A$ will vanish by symmetry in our background.

\subsection{Specifics of the $D0$-background}

Since we consider Fermions below, we will work in terms of the vielbien $e_A^{\hat A}$.  The direction picked out by the $D0$-worldline is clearly special and corresponds to $A=0$.  We will use the symbol $a = \{1, ...9\}$ to indicate one of the directions transverse to the zero-branes.
Thus, the supergravity background is
\begin{eqnarray}
\label{back}
\;\;\;\;&e^\phi=H^{-3/4},\;\;C_0=H^{-1}-1,\;\;\; C_a=0&\;\;\; \cr
&ds^2=e^{\hat A}e^{\hat B}\eta_{\hat A \hat B},\;\ {\rm with} \;\ e^{\hat a}=H^{1/4}\delta^{\hat a}_{b}dx^{b},\ \,  \;\; e^{\hat  0}=H^{-1/4}dt,&\end{eqnarray}
with all other fields vanishing.  The function $H$ is a harmonic function on the nine-dimensional space defined by $x^1,...,x^9$ and\ sourced by the distribution of $D0$-branes.  We will proceed without assuming any particular form for $H$, but for the case of $N_0$ $D0$-branes at the origin $H$ takes the familiar form
$H = 1 + 60 \pi^{2}  \frac{g \ell_s^7  N_0}{r^7}$, where $r^2 = \sum_{a} x^a  x^a.$

The particular form of (\ref{back}) allows a dramatic simplification of the effective action (\ref{Dpt}).
Following the discussion in section 5 of
\cite{MMS2}, it is useful to also impose static gauge $\xi^I = x^I$ for $I = 0,1,2,3,4$ and to impose the
$\kappa$-symmetry gauge
\begin{equation}
\label{kcond}
\bar \psi {1\over 2}(1-\Gamma_{D4})= \bar \psi.
\end{equation}
Thus, from now on we take $\psi$ to be a constrained Fermion satisfying (\ref{kcond}) so that it has only 16 independent components, though the $\Gamma$s are $32\times 32$ matrices.

Finally, at this stage we use our weak coupling approximation to truncate the action by dropping all remaining terms beyond quadratic order in world-volume fields (including interactions between the Fermions and the scalars $X^p$), as such terms give sub-leading contributions in the $g_s \rightarrow 0$ limit.   With this understanding the action (\ref{Dpt}) in the background (\ref{back}) becomes
\begin{eqnarray}
\label{action} &&S_{D4}^{trunc}=S^{(0) \ trunc}_{D4}+S^{(2) \
trunc}_{D4},\ \ {\rm with} \cr &&S_{D4}^{(0) \ trunc}= -T_{D4}\int
d^5x \ \ - T_{D4}  \int d^5x\left\{ \hbox{${1\over
4}$}F^{IJ}F_{IJ} + \hbox{${1\over2}$}\partial^IX^{ p}\partial_IX^{
q}g_{pq} \ +\right . \cr &&\hspace{7.5truecm}\left . -
\hbox{${1\over 8}$} \Theta \tilde
\epsilon^{0ijkl}F_{ij}F_{kl}\right\}\ , \cr &&S^{(2) \
trunc}_{D4}=i T_{D4} \int d^5x \   \bar
\psi\left[\Gamma^I\partial_I-{1\over 8}\partial_I\ln H \
\Gamma^I(1+2\Gamma_{\hat 0}\Gamma^{\hat \varphi})\right] \psi  \ .
\end{eqnarray}
Here we have used indices $I,J$ to denote spacetime directions $\{0,1,2,3,4\}$ on the brane, lower case $i,j$ to denote space directions $\{1,2,3,4\}$ on the brane, and indices $p,q$ to denote directions $\{5,6,7,8,9\}$ transverse to the brane.  Below, we will also use $\hat I, \hat i, \hat p$ to denote the corresponding tangent space directions.  We have also introduced $\Theta=(H^{-1}-1)$, the moduli metric $g_{ pq}=H^{1/2}\delta_{pq}$ and the worldvolume metric $g_{IJ}$
\begin{eqnarray}
g_{IJ}=\left( \begin{array}{cc}
    -H^{-1/2} & 0 \\
    0 & H^{1/2}\delta_{ij}
    \end{array}\right),
\end{eqnarray}
which is used to raise and lower the indices $I,J,i,j$, and the Levi-Civita tensor {\it density} $\tilde \epsilon^{0ijkl}$ whose non-zero entries are $\pm1$.

The supersymmetries of the action (\ref{action}) and their algebra were also derived in \cite{MMS2}.  For completeness, we repeat them here.  They are
\begin{eqnarray}
\label{D4Susy}
&&\delta_{\varepsilon} \psi = \big(\hbox{${1\over4}$}F^{IJ}\Gamma_{IJ}\Gamma^{\hat \varphi}+\hbox{${1\over 2}$}\partial_IX^{ p}\Gamma^I\Gamma_{ p}\big)\varepsilon\ , \cr
&&\delta_{\varepsilon} A_I = i\bar \varepsilon \Gamma_I \Gamma^{\hat \varphi}\psi \ ,\cr
&&\delta_{\varepsilon} X^{ p} =  i\bar \varepsilon \Gamma^{ p}\psi \ ,
\end{eqnarray}
where $\varepsilon=H^{-1/8}\varepsilon^{(0)}$ and  $\varepsilon^{(0)}$ is any a constant spinor satisfying
\begin{eqnarray}
&&\hbox{${1\over 2}$}(1+\Gamma_{\hat  0}\Gamma^{\hat  \varphi})\varepsilon^{(0)}=0 \, \ \  \ {\rm and,}\cr
&&\hbox{${1\over 2}$}(1+\Gamma_{\hat  0 \hat  1 \hat  2 \hat  3 \hat  4}\Gamma^{\hat \varphi})\varepsilon^{(0)}=0.
\end{eqnarray}
Note that the two projectors commute, so that 1/4 of the 32 supersymmetries survives.

>From \cite{MMS2}, the commutator of two such supersymmetry transformations corresponding to $\varepsilon^{1},\varepsilon^{2}$  acting on a bosonic field ($X$ or $A$) is
\begin{equation}
[ \delta_{\varepsilon^{1}},
\delta_{\varepsilon^{2}} ] = \left( -i \bar \varepsilon^{2}  \Gamma^0 \varepsilon^{1} \right)  \partial_0 - Q\left[ i \bar \varepsilon^{2}  \Gamma^0 A_0 \varepsilon^{1} \right],
\end{equation}
where $Q$ is the generator of gauge transformations; i.e. $Q[\Lambda] X = Q[\Lambda] \psi =0$, but $Q[\Lambda] A_i = \partial_i \Lambda$.  In reaching the above form we have used the fact that, since $\Gamma_{\hat  1 \hat 2 \hat 3 \hat 4}
\varepsilon =\varepsilon$, one has $ -i \bar \varepsilon^{2}  \Gamma^I \varepsilon^{1}  =0$.
Note that the factors of $H$ in the first term cancel so that it represents a constant time translation, which is indeed a symmetry of the action (\ref{action}).

We may also use the $\kappa$-symmetry condition (\ref{kcond})  and the identity that $\bar \psi \Gamma^{ABCDE} \psi =0$ for any Majorana spinor $\psi$ to simplify the expression
(\ref{current}) for the current which couples to $C_I$.  The result is
\begin{equation}
\label{current2}
J^I = \frac{T_{D4}}{8 \sqrt{-g}}  \tilde \epsilon^{IJKLM} F_{JK} F_{LM} +
\frac{i T_{D4}}{4 \sqrt{-g}} \partial_J \left( \sqrt{-g} \bar \psi \Gamma^{JI} \Gamma^{\hat \varphi} \psi \right),
\end{equation}
where one is pleased to note that all derivatives transverse to the brane have disappeared.
Note that since the scalars $X^p$ do not appear in (\ref{current2}) and are decoupled from all other fields, they are irrelevant to our calculation and will not appear in any discussion below.

\section{The induced $D0$ charge density}
\label{calc}

We are now nearly ready to compute the expectation value $\langle \rho_{D0}\rangle = \langle J^0 \rangle$ in our background.  In order to properly take into account the deformation of the ground
state, it is useful to compute $\langle \rho_{D0}\rangle$ in the corresponding Euclidean signature background and then to analytically continue back to Lorentz signature.

We find it easiest to keep track of the relevant signs and factors of $i$ by proceeding exactly as stated above; that is, by analytically continuing the {\it background} and making no changes in the {\it coordinates}.  That is, we take the Lorentzian action (\ref{Dpt}) to define a function $S_L(X,F,\psi; b)$, where $b$ is the supergravity background and simply substitute the Euclidean background $b_E$ defined by (\ref{back}) with the replacements
\begin{equation}
\label{bE}
e^{\hat 0} = -i H^{-1/4} dt \ \ {\rm and}  \ \ C = -i (H^{-1} -1) dt.
\end{equation}
In particular, the metric still has the form $ds^2 = e^{\hat A} e^{\hat B} \eta_{\hat A \hat B}$ with $\eta_{\hat A \hat B}$ the Minkowski metric.  The Levi-Civita tensor density $\tilde \epsilon^{IJKLM}$ of course remains $\pm1$ or $0$, but $\sqrt{-g} := det (e)$ changes by the above factor of $-i$.
We also follow the standard convention of introducing another factor of $-i$ in the Euclidean action, which we define for {\it any} background $b$ as $S_E(X,F,\psi;b) = -i S_L(X,F,\psi;b)$.

Evaluating the Euclidean action $S_E(X,F,\psi;b_E)$ on the background $b_E$ of (\ref{bE}) yields:

\begin{eqnarray}
\label{Eaction} &&S_{ED4}^{trunc}=S^{(0)}_{D4}+S^{(2) \
trunc}_{D4},  \cr &&S_{ED4}^{(0)}= T_{D4}\int d^5x + \ \
T_{D4}\int d^5x\left\{ \hbox{${1\over 4}$}F^{IJ}F_{IJ} +
\hbox{${1\over2}$}\partial^IX^{ m}\partial_IX^{ n}g_{ m  n} \
\right . \cr &&\hspace{7.5truecm}\left . - \hbox{${1\over 8}$}
\Theta \tilde \epsilon^{0ijkl}F_{ij}F_{kl}\right\}\ .
\end{eqnarray}

To display the Fermion action it is useful to first clarify our definition of the analytic continuation.  We take $\Gamma^{\hat A}$ to be {\it independent} of the background, with $\Gamma^A$ defined in terms of $e^{\hat A}$ and $\Gamma^{\hat A}$.  Thus, $\Gamma^0$ depends on the background; it is anti-Hermitian in a Lorentzian background and Hermitian in a Euclidean one.  On the other hand, $\Gamma^{\hat 0}$ is always anti-Hermitian.  However, we find it convenient to define $\Gamma_E^{\hat 0} := i \Gamma^{\hat 0}$ and $\Gamma_{E, \hat 0} := -i \Gamma_{\hat 0}$, from which we see that $\Gamma_E^{\hat 0} =  \Gamma_{E, \hat 0}$.  Note that we have:
\begin{eqnarray}
\Gamma_{D4} = i \Gamma_{E\hat 0} \Gamma_{\hat 1}\Gamma_{\hat 2}\Gamma_{\hat 3}\Gamma_{\hat 4} \Gamma^{\hat \varphi}, \ \ \ {\rm and} \\
\Gamma^{\hat \varphi} = -i \Gamma^{E\hat 0} \Gamma^{\hat 1}\Gamma^{\hat 2}\Gamma^{\hat 3}\Gamma^{\hat 4} \Gamma^{\hat 5} \Gamma^{\hat 6}\Gamma^{\hat 7}\Gamma^{\hat 8}\Gamma^{\hat 9}.
\end{eqnarray}
We also introduce $\bar \psi_E = -i \bar \psi.$  With such understandings the quadratic Fermion action is

\begin{equation}
 S^{(2) \
trunc}_{ED4}=- T_{D4}\int d^5x \   \bar
\psi_E\left[\Gamma^I\partial_I-{1\over 8}\partial_i\ln H \
\Gamma^i(1+i 2\Gamma_{E,\hat 0}\Gamma^{\hat \varphi})\right] \psi  \
,
\end{equation}
where $\psi$ continues to satisfy
\begin{equation}
\frac{1}{2}\bar \psi_E (1 - \Gamma_{D4}) = \bar \psi_E.
\end{equation}

Similarly, we define a Euclidean current $J^0_{E}(X,F,\psi;b)$ on
{\it any} background $b$ through $J^0_E(X,F,\psi;b) :=
-iJ^0_L(X,F,\psi;b_E) $.  Evaluating this current on $b_E$ yields
\begin{equation}
\label{Ecurrent} J^0_E =    \frac{T_{D4}}{8H^{3/4}}  \tilde
\epsilon^{0ijkl} F_{ij} F_{kl} + i \frac{ T_{D4}}{4H^{3/4}}
\partial_i \left(H^{3/4} \bar \psi_E \Gamma^{\hat i} \Gamma_E^{\hat0} \Gamma^{\hat
\varphi} \psi \right).
\end{equation}
It will be convenient to restrict attention to weak supergravity
fields so that we may treat the system perturbatively.  This
amounts to the condition $ \delta H \equiv H-1 \ll 1$, so that we
may approximate $\partial_i \ln H \approx
\partial_i H$ and $\Theta = H^{-1} -1 \approx 1- H$.

Because $J^0_E$ contains products of operators at coincident points, the individual terms are likely to be divergent.  Our strategy will be to point-split each term along some displacement $\delta$ in the Euclidean time direction and then add the contributions from each term together, analyzing the limit $\delta \rightarrow 0$.  The purely Bosonic part $J^0_{bE}$ of the current (\ref{Ecurrent}) will be
studied in subsection \ref{bose} below, while the part $J^0_{fE}$ quadratic in Fermions will be studied in subsection \ref{fermi}. We will then collect the terms and study the coincidence limit in subsection
\ref{limit}.  The reader may wonder what happens to these divergences in the trivial background $H=1$.   As we will see below, it turns out that the index and $\Gamma$-matrix structure of (\ref{Eaction}) and (\ref{Ecurrent}) cause both contributions to $J^0_E$ to vanish identically for $H=1$, even at finite point-splitting parameter $\delta.$

\subsection{The Bosonic part of the Euclidean Current}
\label{bose}

Let us now consider the point-split bosonic contribution,
\begin{equation}
\label{Jb} \langle J^0_{bE} (x,y) \rangle =   \frac{T_{D4}}{8}
\tilde \epsilon ^{0ijkl} \langle F_{ij}(x) F_{kl} (y) \rangle (1 + \O(\delta H)) = -
\frac{T_{D4}}{2}  \epsilon^{0ijkl} \partial_{i_x} \partial_{j_y}
\langle A_{k}(x) A_{l} (y) \rangle (1 + \O(\delta H)) ,
\end{equation}
where we have  written this result in terms of the two-point function of the world-volume connection $A_J$ that leads to the field strength $F_{IJ}$.  We have also
explicitly indicated the two arguments $x,y$ of the point-split current.  The subscripts $x,y$ on indices indicate the points at which the corresponding derivatives act.
The two point function $\langle A_{k}(x) A_{l} (y) \rangle$ may be computed from the equation of motion for $F$, which may be written
\begin{equation}
\label{FEOM}
\partial_I F^{IJ} =  - \frac{1}{2}  \tilde \epsilon^{0iJkl} F_{kl}  \partial_i H + O(\delta H^2).
\end{equation}
The two-point function satisfies this same equation, but with an additional delta-function source.

As we will solve the problem perturbatively, we wish to express (\ref{FEOM}) in the form
\begin{equation}
\delta^{JL} \delta^{IK}\partial_I F_{KL} = L^{JI} A_I,
\end{equation}
where $L^{JI}$ is a linear differential operator that is also
linear in $\delta H$.  Since every term in (\ref{FEOM}) contains
two derivatives (which act either on $A^K$ or on $H$), each term
in $L^{JI}$ must contain two derivatives as well (which act either
on $H$ or on the argument of $L^{JI}$). The perturbative solution
for the two-point function will then  be
\begin{equation}
\label{Apert} T_{D4} \langle A_{K}(x) A_{L} (y) \rangle =
G_{KL}(x,y) - \int d^5 z G_{KI}(x,z) L^{IJ} G_{JL}(z,y),
\end{equation}
where $G_{KL}(x,y)$ is the flat-space two-point function (corresponding to $H=1$).

>From (\ref{FEOM}) there are two possible sources of corrections to
the flat-space two-point function $\langle A_{k}(x) A_{l} (y)
\rangle_0$.  The first is from the metric factors used to raise
the indices in $F^{IJ}$ on the left-hand side of (\ref{FEOM}), the
second is  from the explicit source term on the right-hand side.
At lowest order in $\delta H$ the full correction term is the sum
of these two independent sets of corrections.

Let us consider the first set of corrections, working in the Euclidean version of flat-space Lorentz gauge: $\delta^{IJ} \partial_I A_J =0$; i.e., in a gauge that preserves all symmetries and in which $G_{KL} =  (\delta_{KL} - \frac{\partial_K\partial_L}{\partial^2}) G$, where $\partial^2 = \delta^{IJ} \partial_I \partial_J$ is the flat Euclidean Laplacian and $G$ is the scalar Green's function satisfying
\begin{equation}
\partial^2 G(x,y) = - \delta(x,y).
\end{equation}
Consider in particular the contribution of such corrections to the
factor $\partial_i \partial_j \langle A_{k}(x) A_{l} (y) \rangle$
appearing in (\ref{Jb}).  Note that (\ref{Jb}) contracts this with
$\tilde \epsilon^{0ijkl}$, so that we may neglect any terms proportional
to the flat-space metric (on any pair of indices).  Thus,
non-trivial terms can arise only when each index is generated by
the action of a derivative ($\partial_i, \partial_j,\partial_k,
\partial_l$) on one of the Green's functions or on $H$.  Since
$\partial_i,\partial_j$ are explicit derivatives and $L^{JI}$
contains two additional derivatives, there are indeed four
derivatives in each such correction term. However, each of
these four derivatives must act on $G(x,z)$, $G(z,y)$, or $H(z)$.
Thus, some two of these derivatives act on the same function and,
when antisymmetrized by contraction with $\tilde \epsilon^{0ijkl}$, cause
the result to vanish.  Thus, we may neglect all factors of $H$ in
the metric and replace $L^{JI}$ by
\begin{equation}
L_{right}^{J L} =  -  \tilde \epsilon^{0iJkL}   (\partial_i H)
\partial_k.
\end{equation}

Similarly, the anti-symmetry of $\tilde \epsilon^{0ijkl}$ implies that the zero-order contribution to $J^0_{bE}$ vanishes for all $x,y$.  Since we wish to compute
$\tilde \epsilon^{0ijkl} \partial_{i_x} \partial_{j_y} \langle A_{k}(x) A_{l} (y) \rangle$ (i.e., a correlator of field strengths), it is also clear that we may simply replace $G_{IJ}$ by $\delta_{IJ} G$, dropping the longitudinal correction term  $- \frac{\partial_K\partial_L}{\partial^2} G$, as this term will again lead to commutators of coordinate derivatives.
Thus, we have
\begin{eqnarray}
\label{bJresult} \langle J^0_{bE} (x,y)\rangle &=& \frac{1}{2}
\tilde \epsilon^{0ijkl} \tilde \epsilon^{0i'j'k'l'} \delta_{kk'} \delta_{ll'}
\int d^5 z  \partial_i  G(x,z) [\partial_{i'}H(z)]
\partial_{j'} \partial_j G(z,y) \cr &=&   - (\delta^{ik}
\delta^{jl} - \delta^{il} \delta^{jk}) \int d^5 z \partial_i
G(x,z) [\partial_{j}\partial_k H(z)] \partial_l G(z,y),
\end{eqnarray}
where all derivatives are with respect to the $z^i$ coordinates.
We will postpone detailed analysis of the limit $x \rightarrow y$ until after computation of the fermion contribution $J^0_{fE}$, to which we now turn.

\subsection{The Fermionic part of the Euclidean Current}
\label{fermi}

Our approach to the Fermionic contribution $J^0_{fE}$ will proceed in parallel with our calculation of
the bosonic term $J^0_{bE}$ above.  We wish to consider the Fermionic term
\begin{equation}
\label{JfE}
\langle J^0_{fE}(x,y) \rangle =
 i \frac{ T_{D4}}{4} \partial_i \left(H^{3/4} \langle  \bar \psi_E (x) \psi (y) \rangle^\beta_\alpha  (\Gamma^{\hat i}\Gamma_E^{\hat 0} \Gamma^{\hat \varphi})^\alpha_\beta \right),
\end{equation}
where $\partial_i = \partial_{i_x} + \partial_{i_y}$ acts on functions of both $x$ and $y$ and where we have explicitly displayed the spinor indices $\alpha, \beta$.  It will not matter at which point the factor $H^{3/4}$ is evaluated as we will shortly see that to leading order we may replace this factor with $1$.

The two-point function is again determined by the equation of motion, which for the spinor $\psi$ is just
\begin{equation}
\label{pEOM}
\dsl_H \psi = {1\over 8}\partial_I\ln H \ \Gamma^I(1+ i 2\Gamma_{E,\hat 0}\Gamma^{\hat \varphi}) \psi,
\end{equation}
where $\dsl_H = \Gamma^I \partial_I$ and the subscript indicates the implicit dependence on $H$.
Again, we wish to express (\ref{pEOM}) as a linear perturbation of the flat-space result:
\begin{equation}
\dsl \psi = L \psi,
\end{equation}
where $\dsl = \dsl_0 =  \Gamma_E^{\hat 0} \partial_0 + \Gamma_E^{\hat{i}} \partial_i$ and $L$ is linear in $\delta H$.  Here it is useful to introduce the notation
\begin{equation}
\Gamma_E^{\hat I} = \Biggl\{ {{\Gamma_E^{\hat 0} \ \ \ {\rm for } \ I=0} \atop {\Gamma^{\hat j} \ \ \ {\rm for } \ I=j}},
\end{equation}
so that we may write $\dsl_0 = \Gamma^{\hat I}_E \partial_I$.
 The two-point function is then
\begin{equation}
\label{ppert}
2T_{D4}  \langle \bar \psi_E (x) \psi (y) \rangle^\beta_\alpha = G^\beta_\alpha (x,y) - \int d^5 z \ G^\beta_{\gamma}(x,z) L_\sigma^{\gamma} G^{\sigma}_\alpha(z,y),
\end{equation}
where $G_\beta^\alpha (x,y) = ( P\dsl G(x,y))^\alpha_\beta$
where $G(x,y)$ is again the scalar Green's function, the derivatives act on the first argument, and $P = \frac{1- \Gamma_{D4}}{2}$ is the projection
onto spinors satisfying the constraint (\ref{kcond}).  Note that the $2$ on the left-hand side of (\ref{ppert}) is a result of our unconventional normalization of the action for Majorana Fermions.

As in the bosonic case, we may consider two sorts of contributions to $L$: those from the left-hand side of (\ref{pEOM}) and those from the right-hand side.
Contributions from the left-hand side yield $L^{left} =\frac{1}{4}
(1-H)(\Gamma^{\hat{0}}_E \partial_0- \Gamma^{\hat{j}} \partial_j)$. Note that $(L^{left \ \gamma}_\sigma
G_\alpha^\sigma)(z,y) = \frac{1}{4}(1-H)(- \eta^{IJ} \partial_I \partial_J G (1-P)^\gamma_\alpha + 2
\partial_0 \partial_j G (\Gamma^{\hat{0}}_E P \Gamma^{\hat{j}})^\gamma_\alpha).$ As a result, the
corresponding contributions to (\ref{JfE}) involve traces of the matrices $ P \Gamma^{\hat J}_E
\Gamma^{\hat{i}}\Gamma_E^{\hat 0}\Gamma^{\hat \varphi}$ and $ P \Gamma^{\hat J}_E  \Gamma_E^{\hat 0} P
\Gamma^{\hat{k}}  \Gamma^{\hat{i}}\Gamma_E^{\hat 0}\Gamma^{\hat \varphi} = P \Gamma^{\hat J}_E
\Gamma^{\hat{k}}\Gamma_E^{\hat i}\Gamma^{\hat \varphi}$, where we have used $P^2 =P$. But we have
\begin{equation}
\label{comm} (1 - \Gamma_{D4}) \Gamma_E^{\hat J}  \Gamma^{\hat{i}}\Gamma^{\hat I}\Gamma^{\hat \varphi} =
\Gamma^{\hat \varphi}(- 1 - \Gamma_{D4})  \Gamma_E^{\hat J}  \Gamma^{\hat{i}}\Gamma^{\hat I},
\end{equation}
so it is in fact sufficient to average the left- and right-hand sides of (\ref{comm}) and, using
cyclicity of the trace, to compute the trace of $\Gamma_E^{\hat J}  \Gamma_{D4} \Gamma^{\hat \varphi}
\Gamma^{\hat{i}}\Gamma_E^{\hat I}.$ However, this operator anti-commutes with any $\Gamma^k$ for which
$k\neq i,I,J$ and so must have vanishing trace.
Thus it is sufficient to replace $L$ by $L_{right}  =  {1 \over
8}\partial_I\ln H \ \Gamma_E^{\hat I}(1+2i \Gamma_{E, \hat 0}\Gamma^{\hat
\varphi})$.

Similar $\Gamma$-matrix algebra shows that $G_\alpha^\beta (x,y) (\Gamma^{i}\Gamma_E^{\hat 0} \Gamma^{\hat \varphi})^\alpha_\beta =0$ so that the zero-order contribution to $\langle J^0_{fE} \rangle$ vanishes identically at any $x,y$.  The full expectation value is therefore
\begin{eqnarray}
\label{JfEresult}
\langle J^0_{fE}(x,y) \rangle &=&-
\frac{ 1}{64}   Tr[P \Gamma_E^{\hat I}   \Gamma^k(-i +2\Gamma_{E, \hat 0}\Gamma^{\hat \varphi}) P \Gamma_E^{\hat J}  \Gamma^{l}\Gamma_E^{\hat 0} \Gamma^{\hat \varphi}] \cr &\times& (\partial_{l_x} + \partial_{l_y})
  \int d^5 z \ \partial_I G(x,z) [\partial_k H(z)] \partial_J G(z,y) + \O(\delta H^2),
  \end{eqnarray}
where all derivatives inside the integral are performed with respect to $z$.  Note that to this order we may replace each remaining $\Gamma^I$ by its flat-space counterpart.
We find
\begin{equation}
Tr[P\Gamma_E^{\hat I}   \Gamma^k(-i+2\Gamma_{E,\hat 0}\Gamma^{\hat \varphi}) P\Gamma_E^{\hat J}  \Gamma^l \Gamma_E^{\hat 0} \Gamma^{\hat \varphi}]
=  16(- \tilde \epsilon^{0IJkl} - 2 [\delta^{Il}\delta^{kJ} + \delta^{Ik} \delta^{Jl} - \delta^{IJ} \delta^{kl} + 2 \delta^{0I} \delta^{0J} \delta^{kl} ])   + \O(\delta H).
\end{equation}
However, the term involving $\tilde \epsilon^{0IJkl}$ will contribute a term to $\langle J^0_{fE}\rangle$ proportional to the commutator of two derivatives, which of course vanishes.  Thus the fermionic contribution is
\begin{equation}
\label{JfEresult2}
\langle J^0_{fE}(x,y) \rangle =
 \frac{1}{2} (2\delta^{Il}\delta^{kJ}  - \delta^{IJ} \delta^{kl}  + 2\delta^{0I} \delta^{0J} \delta^{kl})
  \int d^5 z \ \partial_I G(x,z) [\partial_l \partial_k H(z)] \partial_J G(z,y) + \O(\delta H^2),
  \end{equation}
  where we have used the fact that $\partial_k \partial_l = \partial_l \partial_k $ to simplify
  the factor involving Kronecker delta's, and we continue with the convention that all derivatives inside the integral are with respect to $z$.  We note that this expression is structurally quite similar
to the bosonic contribution (\ref{bJresult}), except that a different combination of derivatives is involved as well as a different overall coefficient.  In particular, Euclidean time derivatives of $G$ {\it do} appear in the fermion contribution, while they were absent in (\ref{bJresult}).  We also note that, in comparison with the bosonic contribution,  the Fermionic contribution weights the term where $l,k$ are contracted with $I,J$ by an extra factor of two relative to the term where $l,k$ are contracted together.  As we will shortly see below, these features will
prevent the bosonic and fermionic divergences from canceling.

\subsection{The coincidence limit}

\label{limit}

Having obtained the expressions (\ref{bJresult}) and (\ref{JfEresult2}), we now turn to an exploration
of the coincidence limit $x \rightarrow y$.  To this end, it will be convenient to reparametrize the problem
in terms of the average location $\Delta_+^I =  (x^I+y^I)/2$ and the difference $\Delta_-^I = (x^I-y^I)/2$ of the two points.  We will take the separation to be purely in the Euclidean time direction, so that $\Delta_-^i=0$.  We will be most interested in the singular contributions to (\ref{bJresult}) and (\ref{JfEresult2}), which result from
the region where $z$ is close to either $x$ or $y$.  As a result, it is convenient to parametrize $z$ as
$z^I =  \Delta_+^I+ |\Delta_-^0| \eta^I$.

Recall that the explicit form of the scalar Green's function is $G(x,y) = \frac{1}{3 V(S^4) |x-y|^3}$, where $V(S^4)$ is the volume of a unit $S^4$ and $|x-y|$ is the Euclidean length of the 5-vector $x-y$.  Using this result, we may write our results in the form
\begin{equation}
\langle J_E^0(x,y) \rangle =   \frac{-1}{|\Delta_-|^3}
\frac{1}{[V(S^4)]^2} \int d^5 \eta \frac{(\eta^I -
\Delta_-^I/|\Delta_-|)(\eta^J + \Delta_-^J/|\Delta_-|)}{|\eta
-\hat x^0|^5\ \ |\eta + \hat x^0|^5}
 A_{IJ}(\Delta_+ + |\Delta_-^0| \eta),
\end{equation}
where $\hat x^0$ is a unit vector in the positive $x^0$ direction.   Note in particular that $|\Delta_-| = |\Delta_-^0|$, the absolute value of the time component of $\Delta_-$.  In the above expression,
\begin{eqnarray}
A_{IJ}(z) &=&  A^{bose}_{IJ}(z) +A^{fermi}_{IJ}(z) \ \ \ {\rm
with} , \cr A^{bose}_{IJ} (z) &=&   (\delta^{k}_I \delta^{l}_J - \delta_{IJ} \delta^{lk}+
\delta^{lk} \delta_{J0}\delta_{I0})
\partial_k \partial_l H(z), \cr
A^{fermi}_{IJ}(z) &=&
- \frac{1}{2} (2\delta^{l}_I\delta^{k}_J  - \delta_{IJ} \delta^{kl}+ 2 \delta_{0I} \delta_{0J} \delta^{kl}) \partial_k
\partial_l H(z).
\end{eqnarray}

One may now expand $A_{IJ}(z)$ about the point $\Delta_+$ to obtain a power series in $|\Delta_-|$.
Since any odd parity integrand will integrate to zero, only even terms in this expansion will contribute.
Furthermore, terms of order $z^4$ or higher in $A_{IJ}(z)$ will give vanishing contribution in the limit
$\Delta_- \rightarrow 0$.  Thus, the only relevant terms involve $A_{IJ}(\Delta_+)$ and  $\partial_K \partial_L A_{IJ}(\Delta_+)$.  The complete list of relevant integrals is provided in appendix A, and quickly yields the result

\begin{eqnarray}
\label{finalJ} \langle J_{bE}^0(x,y) \rangle &=&  \frac{1}{12}
\frac{V(S^3)}{[V(S^4)]^2} \left( \frac{1}{|\Delta_-|^3}
\partial_\perp ^2 H(\Delta_+) + \frac{1}{3} \frac{1}{|\Delta_-|}
( \partial_\perp ^2)^2 H(\Delta_+) \right) + \O(\Delta_-),
 \cr
\langle J_{fE}^0(x,y) \rangle &=&   - \frac{1}{18}
\frac{V(S^3)}{[V(S^4)]^2}  \frac{1}{|\Delta_-|^3}
\partial_\perp ^2 H(\Delta_+) +  \O(\Delta_-),
\end{eqnarray}
where $\partial_\perp^2 = \delta^{ij}\partial_i \partial_j$ is the Laplacian in the 1,2,3,4 directions.  It is interesting that the $\O(\Delta^{-1})$ Fermion
contribution vanishes.

Finally, we should continue the result back to Lorentzian spacetime.  To do so, let us first compute the Lorentzian current in the Euclidean background $b_E$ defined by (\ref{bE}):
\begin{eqnarray}
J^0_L(b_E) &=& i J^0_E(b_E) = \frac{i}{36}
\frac{V(S^3)}{[V(S^4)]^2|\Delta_-|^3}
\partial_\perp ^2 H(\Delta_+) (1 + \O(\Delta_-^3))\cr
&=&  \frac{1}{36} \frac{V(S^3)}{[V(S^4)]^2|\Delta_-|^3}
\partial_\perp ^2 C^0(b_E) (\Delta_+) (1 + \O(\Delta_-^3)),
\end{eqnarray}
where we have used (\ref{bJresult}) to see that the factor of $H$
above came only from $C^0(b_E) = g^{00}(b_E) C_0(b_E) =
-i(H^{-1}-1) = i \delta H + \O(\delta H^2)$, where the $g^{00}$ results from the contraction of the two Levi-Civita symbols. Thus, we may
analytically continue to a Lorentzian background to find
\begin{equation}
\langle J_L^0 \rangle = \frac{1}{36}
\frac{V(S^3)}{[V(S^4)]^2|\Delta_-|^3}
\partial_\perp ^2 C^0(b_E) (\Delta_+) (1 + \O(\Delta_-^3)).
\end{equation}

\section{Discussion}
\label{disc}

The calculations above find that, when a $D4$-brane probe is placed in the supergravity background generated by $D0$-branes, the expectation value of the point-split $D0$-brane charge density $\langle \rho_{D0}(x,y)\rangle$ is non-zero at leading order.  Furthermore, our low-energy field theory calculation gives a  divergent result in the coincidence limit $x \rightarrow y$.  However, we note that the divergences are proportional to $\partial_\perp^2 H$ or $(\partial_\perp^2)^2 H$.  Thus, as one would expect from charge conservation, they yield zero total induced $D0$-charge when integrated over the $D4$-brane.  We also note that this charge density vanishes in the limit $H \rightarrow 1$ where the $D0$ source is infinitely far away.

Such calculations clearly beg a fully string theoretic calculation.
We reserve such a Ramond-Ramond calculation for future work, but it is natural
to expect the result to merely cut off our field-theoretic
divergence at the string scale, $\Delta_-^0 \sim \ell_s$, yielding
a finite induced Lorentz-signature charge density of the form
\begin{equation}
\label{string} \langle \rho_{D0}(x) \rangle = \frac{1}{36}
\frac{V(S^3)}{[V(S^4)]^2|\Delta_-|^3}
\partial_\perp ^2 H(\Delta_+) (1 + \O(\Delta_-^3)) \sim  a \ell_s^{-3} \partial_\perp ^2 H(x) (1 +
\O(\ell_s^2)),
\end{equation}
where $a$ is an unknown coefficient of order $1$ which one naively expects
to be positive\footnote{\label{AA} Of course, it is possible that this
quantum induced effect could be cancelled by some intrinsic
c-number $\O(g)$ correction to the $D0$ charge density on a
$D4$-brane.  We thank Allen Adams for raising this possibility.}.
In particular, (\ref{string}) indicates a non-vanishing quantum
polarization of the $D4$-brane by the $D0$-background, although
such an effect does not occur classically at any order in
$\alpha'$.

The effect arises because the boson and fermion contributions fail to cancel, though they do have opposite signs.  We note that the sign of the final effect is the natural one expected of a polarizable medium, which follows from the natural tendency of an applied electric field to separate (in this case, virtual) charges.

It also interesting to ask what a term of the form (\ref{string}) would imply for the quantum-corrected low-energy $D4$-brane effective action\footnote{We thank Joe Polchinski for raising this question.}.  That is, we may ask what term in an action $S_{D4}^{qc}$ would, when treated classically, yield an induced charge density of this form.  The charge density is by definition the variation of $S_{D4}^{qc}$ with respect to the background Ramond-Ramond field $C^{(1)}$.  Thus, a charge density linear in the background fields could in principle arise from a quadratic term involving two powers of $C^{(1)}$, or from a term involving one power of $C^{(1)}$ and one power of the metric or dilaton.  However, there are no Lorentz-invariant quadratic couplings of a 1-form to a metric or scalar, so the coupling must be quadratic in $C^{(1)}$.  The Lorentz invariant such term that leads to (\ref{string}) is
\begin{equation}
\label{stringterm} - \frac{a}{4} (2\pi)^4 T_{D4} \int d^5x \left(
g_s\ell_s^2  {\bf F^{(2)}}_{IJ}{\bf F^{(2)}}^{IJ} \right),
\end{equation}
where we have used
$T_{D4} =\frac{1}{(2\pi)^4 \ell_s^5 g_s}$ to write this term using the familiar normalizations of the $D4$-effective action in order to make clear that it does indeed have the form of a first order correction in $g_s$; i.e., a one-loop (annulus) string correction.  Note that ${\bf F^{(2)}}_{IJ}$ is the pull-back of the bulk Ramond-Ramond two form field strength to the brane\footnote{Note that that quantum corrections of our form arise only from corrections to the Green's functions that follow from couplings of bulk fields to world-volume fields in the classical $D4$-effective action.  Since $C^{(1)}$ appears in this action only through its pull-back, corresponding terms induced in the quantum-corrected effective action must also involve only the pull-back of $C^{(1)}$.}.  Thus, we expect the quantum-corrected $D4$ action to contain pull-backs of bulk kinetic terms.

We note, however, that such terms are known from \cite{BBG}  {\it
not} to arise for type II branes at order $g_s^0$ at any order in
$\alpha'$, though Einstein-Hilbert terms on the brane do arise as
$\alpha'$ correction to branes in bosonic string theory
\cite{LCR}.  Returning to the type II context, one may expect
that, in order for terms (\ref{stringterm}) to reside in a
supersymmetric effective action or to follow from a covariant term
in the M5-brane effective action\footnote{We thank Savdeep Sethi
for raising this latter question}, an Einstein-Hilbert term for
the world-volume metric would also be required.  As pointed out in
\cite{DGP}, such a term could have interesting cosmological
implications in braneworld scenarios.  However, this term appears
not to arise \cite{FE} for type II branes\footnote{In particular,
after our original posting of this paper on the arxiv, the author
of \cite{FE} shared with us his unpublished calculations which
explicitly show that the coefficient of the Einstein-Hilbert term
vanishes.  One is tempted to believe that supersymmetry lies
behind the vanishing of this coefficient, but no argument for this
seems to be known.}.    It is not clear to us how this tension is
resolved, though it may be that the quantum polarization term is
cancelled by an explicit $O(g)$ term as suggested above in
footnote \ref{AA}.
\medskip

{\bf Acknowledgments:} We would like to thank Allen Adams, Andrew
Frey, Ken Intrilligator, David Lowe, Luca Martucci, Joe Schechter,
Savdeep Sethi, Mark Srednicki, Amanda Peet, Joe Polchinski, Radu
Roiban, Arkady Tseytlin,  and others whom we have surely forgotten
for a number of useful conversations. We would especially like
to thank Rob Myers, in discussion with whom this project
originally arose and who participated in the early stages of its
development, and Friedel Epple for sharing some of his unpublished calculations. B.C.P. and D.M. were supported in part by NSF grants
PHY0098747, PHY0354978, and PHY0354978 and by funds from the
University of California. PS would like to thank the hospitality
of the university of Barcelona, where part of this work was
completed. PS was partially supported by INFN, MURST and by the
European Commission RTN program HPRN-CT-2000-00131, in association
with the university of Torino.
\appendix

\section{List of Relevant Integrals}

The following is a list of integrals needed to attain the result \ref{finalJ}.

\begin{eqnarray}
&&\int d^5 \eta
\frac{\eta^i \eta^j}{|\eta -\hat x^0|^5\ \ |\eta + \hat x^0|^5} = \frac{V(S^3)}{36} \delta^{ij} \\ \cr
&&\int d^5 \eta
\frac{(\eta^0 - 1)(\eta^0 + 1)}{|\eta -\hat x^0|^5\ \ |\eta + \hat x^0|^5} = - \frac{V(S^3)}{18} \\ \cr
&&\int d^5 \eta
\frac{\eta^i \eta^j (\eta^0 - 1)(\eta^0 + 1)}{|\eta -\hat x^0|^5\ \ |\eta + \hat x^0|^5} = 0 \\ \cr
&&\int d^5 \eta
\frac{\eta^i \eta^j \eta^k \eta^l}{|\eta -\hat x^0|^5\ \ |\eta + \hat x^0|^5} =  \frac{V(S^3)}{54} (\delta^{ij} \delta^{kl} + \delta^{ik} \delta^{jl} + \delta^{il} \delta^{jk})
\end{eqnarray}

\end{document}